\documentclass[%
 aip,
 jmp,%
 amsmath,amssymb,
 reprint,%
]{revtex4-1}

\usepackage{graphicx}
\usepackage{dcolumn}
\usepackage{bm}

\newcommand\curl{\text{\rm curl\,}}

\begin{document}


\title{On $p$-form vortex-lines equations on extended phase space}

\author{M. Fecko}
\email{fecko@fmph.uniba.sk}
\affiliation{Department of Theoretical Physics, Comenius University, Bratislava, Slovakia}



\begin{abstract}
In differential-geometric language, vortex-lines equations on extended phase space of a system may be written as
 $i_{\dot \gamma}d\sigma =0$, where $\sigma$ is a differential 1-form.
 This is the structure, to give a paradigmatic example, of the Hamilton equations.
 Here, we study equations of the same structure, where $\sigma$ is a differential $p$-form.
\end{abstract}

\pacs{02.40.-k, 45.20.Jj, 47.10.Df, 47.32.C-}
\keywords{Hamiltonian mechanics, Nambu mechanics, vortex lines}
\maketitle

\section{\label{sec:intro}Introduction}

In hydrodynamics, \emph{vortex lines} are streamlines of {\it vorticity} vector field
$\boldsymbol \omega$,
which is curl of velocity field $\mathbf v$. In the language of differential forms in $E^3$,
we can incorporate $\mathbf v$ into 1-form
\begin{equation}
\theta =\mathbf v \cdot d\mathbf r
\label{curl1}
\end{equation}
(Refs.~\onlinecite{Arnold1989, Fecko2006}).
Then, for a steady flow, the 2-form $d\theta$ just encodes $\boldsymbol \omega$
\begin{equation}
d\theta = \curl \mathbf v \cdot d\mathbf S \equiv \boldsymbol \omega \cdot d\mathbf S
\label{curl2}
\end{equation}
Now, for any vector field $\mathbf W$, the interior product with $d\theta$ results in
\begin{equation}
i_{\mathbf W}(\boldsymbol \omega \cdot d\mathbf S) = (\boldsymbol \omega \times \mathbf W)\cdot d\mathbf r
\label{curl3}
\end{equation}
Consequently, if we consider a curve $\gamma (t)\leftrightarrow \mathbf r(t)$ in $E^3$, equation
\begin{equation}
i_{\dot \gamma}d\theta = 0
\label{curl4}
\end{equation}
is equivalent to
\begin{equation}
\boldsymbol \omega \times \dot {\mathbf r} = 0
\label{curl5}
\end{equation}
i.e., in each point, the curve is forced to proceed along the vorticity vector field $\boldsymbol \omega$
(its parametrization is irre\-le\-vant).
So, equation (\ref{curl4}) represents the vortex-lines equation in hydrodynamics.

It turns out that we encounter the same type of equation in {\it Hamiltonian mechanics}
(see a wonderful exposition in Ref.~\onlinecite{Arnold1989}).
Consider, first, the {\it extended} phase space of a system. Let local coordinates be $(q^a,p_a,t)$.
Introduce a distinguished 1-form, {\it Poincar\'e-Cartan integral invariant}
\begin{equation}
\sigma = p_adq^a - Hdt
\label{poincartan1}
\end{equation}
Then, {\it Hamilton equations}
\begin{equation}
{\dot q}^a = \frac{\partial H}{\partial p_a}
\hskip 1cm
{\dot p}_a = -\frac{\partial H}{\partial q^a}
\label{hamilton}
\end{equation}
may be succinctly written as
\begin{equation}
i_{\dot \gamma}d\sigma = 0
  \hskip 1cm
  \dot \gamma = {\dot q}^a\partial_{q^a}
              + {\dot p}_a\partial_{p_a}
              + \partial_t
\label{hamileq1}
\end{equation}
Therefore, solutions of Hamilton equations (\ref{hamileq1}) are said to represent,
in analogy with (\ref{curl4}), vortex lines of the form $\sigma$.
(The curve $\gamma$ lives in the {\it extended} phase space and its parametrization is,
 according to (\ref{hamileq1}), arbitrary. However, its projection onto the phase space itself,
$\hat \gamma \leftrightarrow (q^a(t),p_a(t))$,
becomes uniquely parametrized in terms of the {\it time} variable $t$.)

Finally, as Takhtajan showed in Ref.~\onlinecite{Takhtajan1994}
(see also Eq. (10) in Ref.~\onlinecite{Fecko1992}),
we also encounter very similar equation in {\it Nambu mechanics}, Ref.~\onlinecite{Nambu1973}.
Consider first Nambu {\it extended} phase space, a space with coordinates
$(x,y,z,t)$. Introduce, following Ref.~\onlinecite{Takhtajan1994},
a 2-form ("ge\-ne\-ra\-li\-zed Poincar\'e-Cartan integral invariant")
\begin{equation}
\hat \sigma := xdy \wedge dz - H_1dH_2 \wedge dt
\label{poincartan2}
\end{equation}
Then one easily checks that the "vortex-lines type" equation
\begin{equation}
i_{\dot \gamma}d\hat \sigma = 0
 \hskip 1cm
  \dot \gamma = {\dot x}\partial_x
              + {\dot y}\partial_y
              + {\dot z}\partial_z
              + \partial_t
\label{nambueq4}
\end{equation}
reproduces {\it Nambu equations}
\begin{equation}
\dot x_i = \epsilon_{ijk} \frac{\partial H_1}{\partial x_j}
           \frac{\partial H_2}{\partial x_k}
\label{nambueq2}
\end{equation}
or, in vector notation,
\begin{equation}
\dot {\mathbf r} = \boldsymbol \nabla H_1 \times \boldsymbol \nabla H_2
\label{nambueq3}
\end{equation}
Note, however, that $\hat \sigma$ is a {\it two}-form rather than a one-form, now.

And, as is also mentioned in Ref.~\onlinecite{Takhtajan1994}, the procedure even works for the $n$-dimensional
version of the Nambu mechanics (Refs.~\onlinecite{Takhtajan1994, Nambu1973}). One should simply write
a straightforward ge\-ne\-ralization of (\ref{poincartan2}),
\begin{equation}
\hat \sigma := x^1dx^2 \wedge \dots \wedge dx^n - H_1dH_2 \wedge \dots \wedge dH_{n-1}\wedge dt
\label{poincartan3}
\end{equation}
and then
\begin{equation}
i_{\dot \gamma}d\hat \sigma = 0
 \hskip 1cm
  \dot \gamma = {\dot x^1}\partial_1
              + \dots
              + {\dot x^n}\partial_n
              + \partial_t
\label{nambueq5}
\end{equation}
instead of (\ref{poincartan2}) and (\ref{nambueq4}), in order to obtain the corresponding Nambu equations
\begin{equation}
\dot x_i = \epsilon_{ijk\dots l} \frac{\partial H_1}{\partial x_j}
                                \frac{\partial H_2}{\partial x_k}
                                \dots
                                \frac{\partial H_{n-1}}{\partial x_l}
\label{nambueq6}
\end{equation}
($\hat \sigma$ is an $(n-1)$-form, now).

So, {\it vortex-lines} type of equation on {\it extended} phase space turns out to play central role in both
Hamiltonian mechanics (where $\sigma$ is $1$-form) as well as Nambu mechanics (where $\sigma$ is $(n-1)$-form).
Therefore, it might be interesting to have a look at the equation in general.
Are there any interesting ''forgotten'' (or, possibly, well-known) equations for $p$ (the degree of the form $\sigma$)
\emph{somewhere between} $p=1$ and $p=n-1$?

\section{\label{sec:decompform}Decomposed form of the equation}

Consider an extended phase space $\mathbb R \times M$ where $M$, the phase space itself, is an $n$-dimensional manifold
with local coordinates $x^i,i=1,\dots ,n$ and $\mathbb R$ represents the time axis (with coordinate $t$).
Vectors tangent to the factor $M$ are called {\it spatial}. Then, if the trajectory of a system is given by a curve
$\gamma (t)$ on $\mathbb R \times M$, its velocity (tangent) vector may be written as
\begin{equation}
\dot \gamma = \partial_t + v
\label{velocity1}
\end{equation}
where $v={\dot x^i}\partial_i$ is spatial, see examples (\ref{hamileq1}), (\ref{nambueq4}) and (\ref{nambueq5}).
The time component is standardly normalized by $\langle dt,\dot \gamma \rangle =1$.
Let $\sigma$ be a $p$-form on $\mathbb R \times M$.
Then, the main object of our interest will be the {\it vortex-lines equation}
\begin{equation}
i_{\dot \gamma}d \sigma = 0
 \hskip 1cm
 \sigma \ \text{is $p$-form}
\label{vortexlines1}
\end{equation}
There is a simple machinery for decomposing differential forms on product manifold
$\mathbb R \times M$ (see Appendix \ref{appdecomp}).
In particular, the $p$-form $\sigma$ may be decomposed as
\begin{equation}
\sigma = dt \wedge \hat s + \hat r
\label{decompoofsigma}
\end{equation}
and for $d\sigma$ we get
\begin{eqnarray} \label{decompoofdsigma1}
  d\sigma &=& dt \wedge (-\hat d \hat s + \partial_t \hat r ) + \hat d \hat r     \\
                 \label{decompoofdsigma2}
          &\equiv& dt \wedge \hat S + \hat R
\end{eqnarray}
Then, taking into account (\ref{velocity1}), we get from (\ref{vortexlines1})
the following {\it two} equations
\begin{equation}
i_v \hat R = -\hat S
 \hskip 1cm
i_v \hat S = 0
\label{twoequations}
\end{equation}
Note, however, that the second one is a simple consequence of the first one (since $i_vi_v=0$).
Therefore, the following (single) equation
\begin{equation}
i_v \hat R = -\hat S
\label{spatialequation1}
\end{equation}
or, in more detail
\begin{equation}
i_v (\hat d \hat r) = \hat d \hat s - \partial_t \hat r
\label{spatialequation2}
\end{equation}
is {\it equivalent} to (\ref{vortexlines1}). The two versions, (\ref{vortexlines1}) and (\ref{spatialequation1}),
express the same content in two different
languages. Namely, equation (\ref{vortexlines1}) uses the language of "complete"
extended phase space $\mathbb R \times M$, whereas (\ref{spatialequation1})
says everything in terms of {\it spatial} objects
(and operations) $\hat r,\hat s,\hat R,\hat S,\hat d,v$, i.e. via "time-dependent objects on $M$".

\section{\label{sec:liouville}Integral invariants, Liouville theorem}

There are features shared by solutions of (\ref{vortexlines1}) \emph{regardless of concrete} $p$,
resulting from the structure of the equations alone.
In particular, the structure alone guarantees existence of a series of \emph{integral invariants}.

First, integral of $\sigma$ over a $p$-\emph{cycle} (closed $p$-dimensional surface) in
$\mathbb R \times M$ is, according to Cartan (see Ref.~\onlinecite{Cartan1922}), \emph{relative} integral invariant.
This means that
\begin{equation}
 \oint_{c_0}\sigma  =  \oint_{c_1}\sigma
\label{relatinvar1}
\end{equation}
holds for any two $p$-cycles $c_0$ and $c_1$, encircling the same tube of solutions.
(For the proof, see Appendix \ref{appproofintinv}.)

Then, integral of $d\sigma$ over an \emph{arbitrary} $(p+1)$-\emph{chain}
($(p+1)$-dimensional surface, not necessarily a cycle) in $\mathbb R \times M$
is \emph{absolute} integral invariant
(meaning there is no restriction to integrate over cycles).

In the same way one easily checks that relative invariants emerge from integrating
\begin{equation}
 \sigma, \ \ \sigma \wedge d\sigma, \ \ \sigma \wedge d\sigma \wedge d\sigma, \ \dots
\label{relatinvar2}
\end{equation}
over appropriate cycles and for corresponding absolute invariants we are to integrate
\begin{equation}
 d\sigma, \ \ d\sigma \wedge d\sigma, \ \ d\sigma \wedge d\sigma \wedge d\sigma, \ \dots
\label{absolinvar1}
\end{equation}
over appropriate chains.

We can choose the chains (in particular, cycles) to be \emph{spatial},
i.e.lying in hypersurfaces of \emph{constant time}.
Then we get original \emph{Poincar\'e integral invariants}.
In their explicit expressions one can restrict the forms under integral sign to their
\emph{spatial} parts (since $dt$ vanishes on the hypersurfaces).
Therefore, we can write Poincar\'e invariants as integrals of the forms
\begin{equation}
 \hat r, \ \ \hat r \wedge \hat d\hat r, \ \ \hat r \wedge \hat d\hat r \wedge \hat d\hat r, \ \dots
\label{relatinvar3}
\end{equation}
(relative) and
\begin{equation}
 \hat d\hat r, \ \ \hat d\hat r \wedge \hat d\hat r, \ \ \hat d\hat r \wedge \hat d\hat r \wedge \hat d\hat r, \ \dots
\label{absolinvar2}
\end{equation}
(absolute).

If the degree of some of the forms in (\ref{absolinvar2}) matches the dimension of the phase space $M$,
we can use it as \emph{the volume} form on $M$. The statement concerning the corresponding integral
invariant is then nothing but the \emph{Liouville theorem}, expressing \emph{invariance of volume}
of arbitrary domain on $M$ under time development.

\section{\label{sec:examples}Examples}

Here we just summarize examples mentioned in Section \ref{sec:intro} in terms of objects
introduced in Section \ref{sec:decompform} and also provide an
example, where the scheme does not work properly.

\noindent
{\bf Example 1.} {\it Hamiltonian mechanics}. Here
\begin{eqnarray} \label{hamil1}
  \hat s &=& -H \hskip 1cm \hat r = p_adq^a     \\
                 \label{hamil2}
  \text{and} \hskip .5cm
  \hat S &=& \hat d H \hskip 1cm \hat R = dp_a\wedge dq^a
\end{eqnarray}
Therefore, equation (\ref{spatialequation1}) says
\begin{equation}
 i_v(dp_a\wedge dq^a) = -\hat d H
\label{hamil3}
\end{equation}
which leads, taking into account
\begin{equation}
 v = {\dot q}^a\partial_{q^a} + {\dot p}_a\partial_{p_a}
\label{hamil4}
\end{equation}
to Hamilton equations (\ref{hamilton}).
There is, in general, a series of relative integral invariants of type (\ref{relatinvar3})
\begin{equation}
 p_adq^a, \ \ p_adq^a \wedge dp_b\wedge dq^b, 
 \ \dots
\label{relatinvarhamil}
\end{equation}
as well as the corresponding number of absolute invariants of type (\ref{absolinvar2})
\begin{equation}
 dp_a\wedge dq^a, \ \ dp_a\wedge dq^a \wedge dp_b\wedge dq^b, 
 \ \dots,
\label{absolinvarhamil}
\end{equation}
the last one providing the (standard) volume form $d^mqd^mp$ on $M$, $n=2m$, used in Liouville theorem.

\noindent
{\bf Example 2.} {\it Basic Nambu mechanics}. Here
\begin{eqnarray} \label{nambu1a}
  \hat s &=& H_1\hat d H_2 \hskip 1.3cm \hat r = xdy\wedge dz     \\
                 \label{nambu2a}
  \hat S &=& -\hat dH_1 \wedge \hat d H_2 \hskip .4cm \hat R = dx\wedge dy\wedge dz
\end{eqnarray}
Therefore, equation (\ref{spatialequation1}) says
\begin{equation}
 i_v(dx\wedge dy\wedge dz) = \hat dH_1 \wedge \hat d H_2
\label{nambu3a}
\end{equation}
which leads, taking into account
\begin{equation}
 v = {\dot x}\partial_x + {\dot y}\partial_y + {\dot z}\partial_z
\label{nambu4a}
\end{equation}
to Nambu equations (\ref{nambueq2}) or (\ref{nambueq3}).

\noindent
{\bf Example 3.} {\it $n$-dimensional Nambu mechanics}. Here
\begin{eqnarray} \label{nambu1b}
  \hat s &=& H_1\hat d H_2 \wedge \dots \wedge \hat dH_{n-1}      \\
                 \label{nambu2b}
  \hat S &=& -\hat dH_1\wedge \hat d H_2 \wedge \dots \wedge \hat dH_{n-1}      \\
                 \label{nambu3b}
  \hat r &=& x^1dx^2 \wedge \dots \wedge dx^n      \\
                 \label{nambu4b}
  \hat R &=& dx^1\wedge dx^2 \wedge \dots \wedge dx^n
\end{eqnarray}
Therefore, equation (\ref{spatialequation1}) says
\begin{equation}
 i_v(dx^1\wedge dx^2 \wedge \dots \wedge dx^n) = \hat dH_1\wedge \hat d H_2 \wedge \dots \wedge \hat dH_{n-1}
\label{nambu5b}
\end{equation}
which leads, taking into account
\begin{equation}
 v = {\dot x^1}\partial_1
              + \dots
              + {\dot x^n}\partial_n
\label{nambu6b}
\end{equation}
to Nambu equations (\ref{nambueq6}).
There is, for general $n$, just a single relative invariant as well as a single absolute invariant
of type (\ref{relatinvar3}) and (\ref{absolinvar2}),
in both cases the first one in the series
(the degrees of remaining forms are too high to accomodate the forms on $M$).
The absolute invariant, integral of $dx^1\wedge dx^2 \wedge \dots \wedge dx^n$, leads to Liouville theorem.

\noindent
{\bf Example 4.} {\it The case $\dim M=3$, $\deg \sigma = 1$}. This should serve as an elementary
                  \emph{counter}example which demonstrates why
                 one has to choose $\sigma$ in (\ref{vortexlines1}) with due caution.
                 So, consider phase space with coordinates $(x,y,z)\equiv (q,p,z)$,
just like in Example 2., but now take
\begin{eqnarray} \label{badhamil1}
  \hat s &=& -H(q,p,z,t) \hskip 1cm \hat r = pdq     \\
                 \label{badhamil2}
  \hat S &=& \hat dH \hskip 2.4cm \hat R = dp\wedge dq
\end{eqnarray}
so that
\begin{equation}
 \sigma = pdq - Hdt
\label{badhamil3}
\end{equation}
is {\it one}-form, now, rather than two-form.
Then, equation (\ref{spatialequation1}) says
\begin{equation}
 i_v(dp\wedge dq) = -\hat dH
\label{badhamil4}
\end{equation}
which leads, taking into account
\begin{eqnarray} \label{badhamil5}
        v &=& {\dot q}\partial_q + {\dot p}\partial_p + {\dot z}\partial_z  \\
                 \label{badhamil6}
  \hat dH &=& (\partial_qH)dq +(\partial_pH)dp+(\partial_zH)dz
\end{eqnarray}
to the following system of equations
\begin{equation}
  \dot q = \frac{\partial H}{\partial p}
           \hskip 1cm
  \dot p = -\frac{\partial H}{\partial q}
           \hskip 1cm
       0 = -\frac{\partial H}{\partial z}
\label{badhamil7}
\end{equation}
Clearly, we can not determine $(q(t),p(t),z(t))$ from (\ref{badhamil7}), since there is no equation
containing $\dot z$.
\vskip .2cm
\noindent
[The last equation is a {\it constraint} on $H$ rather than a~differential equation. It excludes
 $z$-dependance of the Hamiltonian $H$. If it {\it is} fulfilled, there is already no contradiction
 in (\ref{badhamil7}). Now, we {\it can} determine $(q(t),p(t))$, but still we have no information
 about $z(t)$.]
\vskip .2cm

 So, we {\it cannot} use forms (\ref{badhamil1}), built into (\ref{decompoofsigma}),
 to produce a {\it reasonable} vortex-lines equation (\ref{vortexlines1}). The problem lies in
 {\it ranks} of forms concerned.
 In particular, the form $\hat R$ has rank (only) 2 so that in $v\mapsto i_v\hat R$
 one dimension is lost. That's why equation (\ref{spatialequation1}) does not fix $v$ uniquely.
 Therefore, one should take care of ranks, in general. This is done in more detail in the next section.

\section{\label{sec:when}When it does work}

Equation (\ref{spatialequation1}) may be regarded as a linear inhomogeneous system of the structure
\begin{equation}
  Av = a
\label{linear1}
\end{equation}
for unknown vector $v$, whereas the linear operator $A$ and the vector $a$ are given.
If (\ref{vortexlines1}) is to represent a reasonable
system of first-order differential equations (like (\ref{hamilton}), (\ref{nambueq2}) or (\ref{nambueq6})
and \emph{un}like (\ref{badhamil7})), the system (\ref{linear1}) has to possess unique solution $v$ for
any given $a$.
In more detail, the map $v\mapsto i_v\hat R$ (see the l.h.s. of (\ref{spatialequation1})) has to be {\it injective},
otherwise a part of $v$ disappears, i.e. we miss some dotted coordinates in the resulting system of equations.
But it also has to be {\it surjective}, otherwise, for some choice of $\hat S$ on the r.h.s. of (\ref{spatialequation1}),
we get contradictory equation.
So, the map $v\mapsto i_v\hat R$ should be a {\it linear isomorphism} of the ($n$-dimensional)
linear space of all vectors $v$ to the ($\binom np$-dimensional) space of all $p$-forms $\hat S$
in $n$-dimensional space. This needs, first, the equality of the dimensions
\begin{equation}
  n=\binom np
\label{equality}
\end{equation}
Assuming $n$ being given, we have {\it just two} solutions for $p$, the degree of $\sigma$:
\begin{equation}
  p=1 \hskip 1cm \text{or} \hskip 1cm p=n-1
\label{twosolutions1}
\end{equation}
Under these conditions, linear isomorphism is possible. In order the possibility be materialized,
the rank of the map $v\mapsto i_v\hat R$, which is, by definition, the rank of the form $\hat R$,
is to be maximal, i.e. $n$.

So, we can summarize the conditions under which the equation (\ref{vortexlines1}) represents a reasonable
system of first-order differential equations as follows:
\begin{eqnarray} \label{twosolutions2}
  &\hat R& \ \text{is \ exact $2$-form or $n$-form}      \\
                 \label{twosolutions3}
   \text{rank of} \ \ &\hat R& \ \text{is \ $n$}
\end{eqnarray}
(exactness of $\hat R$ follows from (\ref{decompoofdsigma1}) and (\ref{decompoofdsigma2})).

For the second possibility in (\ref{twosolutions2}), the additional requirement (\ref{twosolutions3}) is void.
Indeed, {\it each} (non-zero) $n$-form has automatically rank $n$
(in $n$-dimensional space; more generally, a $p$-form has rank at least $p$, see e.g. Ref.~\onlinecite{Fecko2006}).
This possibility is realized in Nambu mechanics (see (\ref{nambu4b})). So, in Nambu mechanics, there is no concern
about the rank of $\hat R$, since it is automatically correct. Similarly, there is no restriction
on the dimension $n$ of the phase space; it is arbitrary.

For the first possibility in (\ref{twosolutions2}), on the contrary, the additional requirement
(\ref{twosolutions3}) is non-trivial. In general, a~closed 2-form can only have, due to the classical result of Darboux,
\emph{even} rank: 2,4, \dots, $n$ or $n-1$ depending on whether $n$ is even or odd. So, necessarily, \emph{our} $n$
is to be even, $n=2m$, and the rank of $\hat R$ is to be $2m$. In Darboux coordinates, then, $\hat R$ takes the form
(\ref{hamil2}). This possibility is realized in standard Hamiltonian mechanics (see (\ref{hamil2})).

\section{\label{sec:conclusions}Conclusions}

Closer inspection of \emph{ranks} of relevant forms (and corresponding mappings) reveals,
that {\it Hamiltonian} and {\it Nambu} equations actually represent the
{\it only meaningful} realizations of the vortex-lines equation (\ref{vortexlines1}) on extended phase space.
There are, to answer the question addressed at the end of section \ref{sec:intro},
\emph{no} ''forgotten'' equations of this type, already.


\appendix

\section{\label{appdecomp}Decomposition of forms}
\setcounter{equation}{0}
On $\mathbb R \times M$, a $p$-form $\alpha$ may be uniquely decomposed as
\begin{equation}
\alpha = dt \wedge \hat s + \hat r
\label{decomposition1}
\end{equation}
where both $\hat s$ and $\hat r$ are {\it spatial},
i.e. they do not contain the factor $dt$ in its coordinate presentation
(here, we assume adapted coordinates, $t$ on $\mathbb R$ and some $x^i$ on $M$).
Simply, after writing the form in coordinates, one groups together all terms
 which do contain $dt$ once and, similarly, all terms which do not contain $dt$ at all.
 Note, however, that $t$ still can enter {\it components} of any (even spatial) form.
 Therefore, when performing exterior derivative $d$ of a {\it spatial} form, say $\hat r$,
 there is a part, $\hat d \hat r$, which does not take into account the $t$-dependance
 of the components (if any; as if it was performed just on $M$), plus a part which,
 on the contrary, only operates on the $t$ variable. Putting both parts together, we have
 \begin{equation}
 d\hat r = dt \wedge \partial_t \hat r + \hat d \hat r
 \hskip 1cm
 \partial_t \hat r \equiv \mathcal L_{\partial_t}\hat r
\label{donspatial}
\end{equation}
Then, for a general form (\ref{decomposition1}), we get
 \begin{equation}
 d\alpha = dt \wedge (-\hat d \hat s + \partial_t \hat r ) + \hat d \hat r
\label{dongeneral}
\end{equation}

\section{\label{appproofintinv}A proof of (\ref{relatinvar1})}
\setcounter{equation}{0}
The proof is amazingly simple (see Ref.~\onlinecite{Arnold1989}).
 Consider integral of $d\sigma$ over the $(p+1)$-chain $\Sigma$ given by the family of trajectories
 (solutions) connecting $c_0$ and $c_1$ (so that $\partial \Sigma = c_0 - c_1$). Then
 $$
\begin{array} {rcl}
  \int_{\Sigma} d\sigma
       &\overset{1.} {=}& \int_{\partial \Sigma} \sigma = \oint_{c_0} \sigma - \oint_{c_1} \sigma \\
       &\overset{2.} {=}& 0
\end{array}
$$
The second line (zero) comes from observation, that $\dot \gamma$ is tangent to $\Sigma$,
so that integral of $d\sigma$ over $\Sigma$ consists of infinitesimal contributions proportional
to $d\sigma (\dot \gamma,\dots)$, vanishing because of (\ref{vortexlines1}).

\nocite{*}
\bibliography{vortex_revtex41}

\end{document}